\documentclass[conference]{IEEEtran}
\IEEEoverridecommandlockouts
\usepackage{cite}
\usepackage{amsmath,amssymb,amsfonts}
\usepackage{multirow}
\usepackage{algorithmic}
\usepackage{booktabs}
\usepackage{graphicx}
\usepackage{textcomp}
\usepackage{xcolor}
\def\BibTeX{{\rm B\kern-.05em{\sc i\kern-.025em b}\kern-.08em
    T\kern-.1667em\lower.7ex\hbox{E}\kern-.125emX}}
\begin{document}

\title{EN-T: Optimizing Tensor Computing Engines Performance via Encoder-Based Methodology\\
% {\footnotesize \textsuperscript{*}Note: Sub-titles are not captured in Xplore and
% should not be used}
% \thanks{Identify applicable funding agency here. If none, delete this.}
}

\author{\IEEEauthorblockN{Qizhe Wu*, Yuchen Gui*, Zhichen Zeng*, Xiaotian Wang*, Huawen Liang*, Xi Jin*}
\IEEEauthorblockA{\textit{Department of physics, University of Science and Technology of China, Hefei, China *} \\
Email: \{wqz1998, guiyuchen, zhichenzeng, wxtdsg, lhw233\}@mail.ustc.edu.cn}
% \and
% \IEEEauthorblockN{Yuchen Gui}
% \IEEEauthorblockA{\textit{University of Science and Technology of China} \\
% Hefei, China \\
% guiyuchen@mail.ustc.edu.cn}
% \and
% \IEEEauthorblockN{Zhichen Zeng}
% \IEEEauthorblockA{\textit{University of Science and Technology of China} \\
% Hefei, China \\
% zhichenzeng@mail.ustc.edu.cn}

% \IEEEauthorblockN{4\textsuperscript{th} Given Name Surname}
% \IEEEauthorblockA{\textit{dept. name of organization (of Aff.)} \\
% \textit{name of organization (of Aff.)}\\
% City, Country \\
% email address or ORCID}
% \and
% \IEEEauthorblockN{5\textsuperscript{th} Given Name Surname}
% \IEEEauthorblockA{\textit{dept. name of organization (of Aff.)} \\
% \textit{name of organization (of Aff.)}\\
% City, Country \\
% email address or ORCID}
% \and
% \IEEEauthorblockN{6\textsuperscript{th} Given Name Surname}
% \IEEEauthorblockA{\textit{dept. name of organization (of Aff.)} \\
% \textit{name of organization (of Aff.)}\\
% City, Country \\
% email address or ORCID}
}

\maketitle

\begin{abstract}
Tensor computations, with matrix multiplication being the primary operation, serve as the fundamental basis for data analysis, physics, machine learning, and deep learning. As the scale and complexity of data continue to grow rapidly, the demand for tensor computations has also increased significantly. To meet this demand, several research institutions have started developing dedicated hardware for tensor computations. To further improve the computational performance of tensor process units, we have reexamined the issue of computation reuse that was previously overlooked in existing architectures. As a result, we propose a novel EN-T architecture that can reduce chip area and power consumption. Furthermore, our method is compatible with existing tensor processing units. We evaluated our method on prevalent microarchitectures, the results demonstrate an average improvement in area efficiency of 8.7\%, 12.2\%, and 11.0\% for tensor computing units at computational scales of 256 GOPS, 1 TOPS, and 4 TOPS, respectively. Similarly, there were energy efficiency enhancements of 13.0\%, 17.5\%, and 15.5\%.
\end{abstract}


%%%%%%%%图下面加标注(a)(b)(c);加参考文献

\section{Introduction}

AI applications, including search engines, generative AI based on Large Language Models (LLMs), and tools for self-driving cars, have become widespread in everyday life. Many of these AI applications rely on large-scale tensor computations, leading to specialized hardware development for these calculations by researchers. Include NVIDIA's incorporation of Tensor Cores \cite{nvidia} in their Volta, Turing, Ampere, and the latest Hopper architectures; Google's Tensor Processing Unit (TPU)\cite{googletpu}; Graphcore's MK series Intelligence Processing Unit (IPU)\cite{ipu}; SambaNova's SN\cite{sambanova} series deep learning processors; HUAWEI's Ascend\cite{ascend}; and Cambricon's Machine Learning Unit (MLU) \cite{mlu}. For mobile devices, the SnapDragon 8gen3,
MediaTek Dimensity 9300, and Apple A17 have all integrated neural processing units to enhance the energy efficiency of generative AI in mobile SoCs.

\begin{figure} [htbp]
  \flushleft 
  \includegraphics[scale=0.20]{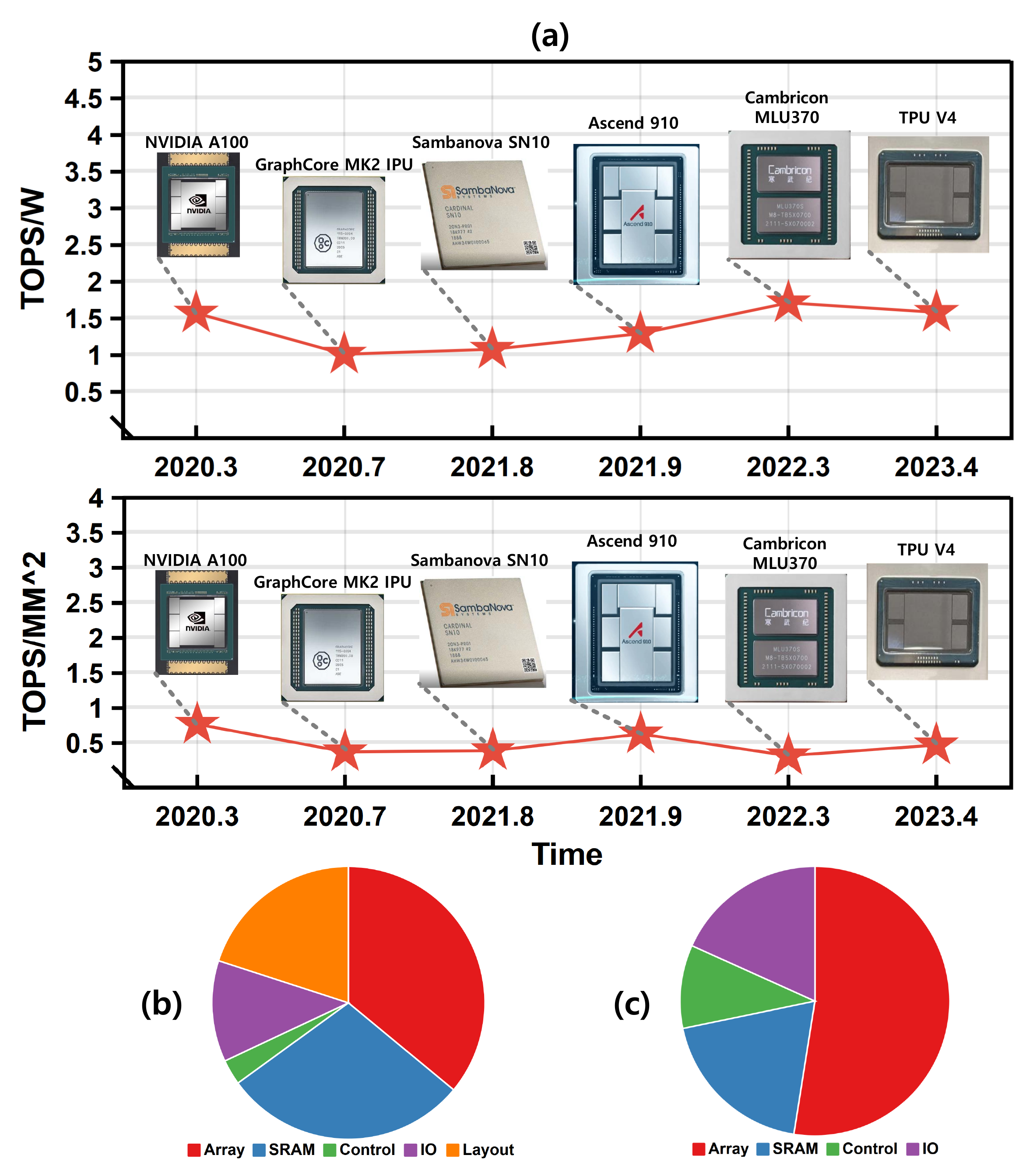}
     \vspace{-0.2cm}
  \caption{(a) Energy efficiency and area efficiency of mainstream 7nm AI accelerators. Area (b) and power (c) breakdown of TPU die.}
  \label{intr_1}
  \vspace{-0.55cm}
\end{figure}

\begin{figure*} [htbp]
  \centering 
  \includegraphics[scale=0.375]{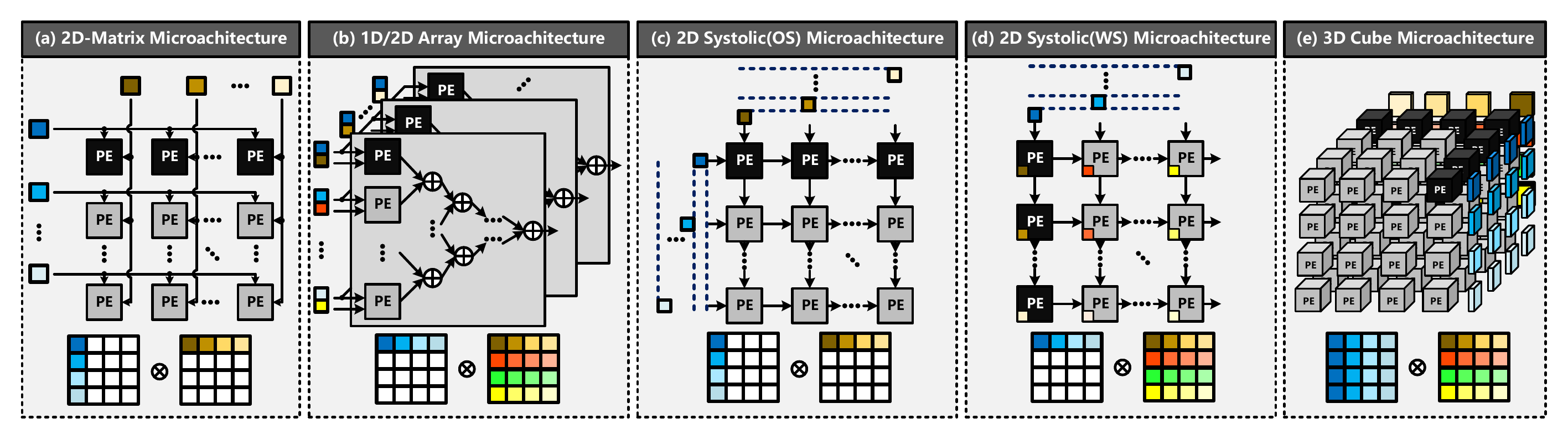}
     \vspace{-0.4cm}
  \caption{Mainstream microarchitectures of Tensor Computing Units in recent years.}
  \label{architecture}
    \vspace{-0.55cm}
\end{figure*}

Fig.\ref{intr_1}(a) shows the INT8 on-chip die performance of 7nm AI processors that have been successfully commercialized in recent years. The rate of performance improvement is gradually decelerating and approaching a stable state. Typical microarchitectures include the 2D Matrix (Fig.\ref{architecture}(a)) of Cambricon's DianNao\cite{diannao} and the 1D/2D Array (Fig.\ref{architecture}(b)) of Cambricon's DaDianNao \cite{dadiannao}, the Systolic Array (Output Stationary (OS) and Weight Stationary (WS)) \cite{msd} 
(Fig.\ref{architecture}(c-d)) of TPU and the speculative Tesla FSD\cite{fsd}, as well as the 3D Cube (Fig.\ref{architecture}(e)) of Ascend \cite{ascend} and NVIDIA \cite{a100}. The primary distinction among these architectures lies in the optimization and alteration of the data flow paths and interconnect topologies within the multiplier arrays, aiming to enhance data reuse for matrix multiplication within the Tensor Computing Units (TCUs). Although some of these architectures were proposed many years ago, they continue to be widely applied in commercial AI processors and academic AI accelerators. Fig.\ref{intr_1}(b)(c) presents the floor plan of the TPU die\cite{googletpu}, showing the area (Fig.\ref{intr_1}(b)) and power (Fig.\ref{intr_1}(c)) consumption distribution. The TCUs, SRAM, and layout wiring occupy 85\% of the die area, with the TCUs (including the multiplier arrays, accumulators, and pipeline registers) accounting for the highest proportion of the area. In terms of power consumption, the TCUs are the primary contributors to the on-chip power. Given the critical role of the TCUs in providing computational power for AI accelerators, optimizing the architecture of these units is particularly crucial for further performance enhancements. 

Our contribution is as follows: (1) We conducted a comprehensive exploration of potential computation reuse in tensor calculations in existing TCUs. To address this, we propose a novel computational paradigm and architecture called EN-T architecture, and developed a novel data encoding representation. This approach offers high versatility and can be seamlessly integrated into existing TCUs to minimize chip area and power; (2)We have implemented our design in RTL using the SMIC 40nm technology in prevalent TCUs in Fig.\ref{architecture}(a$\sim$e). Our results demonstrate an average area efficiency improvement of 8.7\%, 12.2\%, and 11.0\% for TCUs at computational scales of 256 GOPS, 1 TOPS, and 4 TOPS, respectively. Similarly, energy efficiency enhancements of 13.0\%, 17.5\%, and 15.5\%.

\section{Methodology}
\subsection{EN-T Architecture}
\begin{figure}[htbp]\centering\includegraphics[scale=0.4]{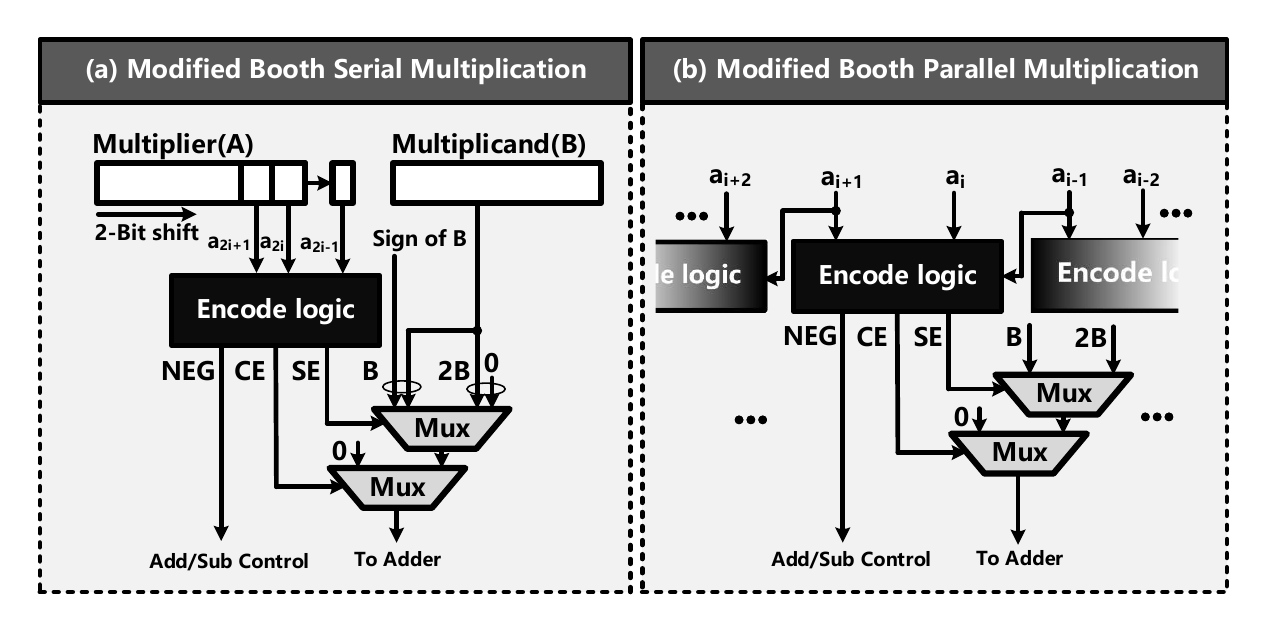}
   \vspace{-0.4cm}
\caption{Modified Booth multiplier.}
\label{booth}
   \vspace{-0.55cm}
\end{figure}

In the design of multipliers in modern computer systems, the multiplication ($A\times B$) involves three key steps. First, partial products of the multiplier $B$ are generated; second, all the partial products are compressed to produce the final row of sums and carries; lastly, a full adder is used to accumulate the sums and carries to obtain the product result. Many designs adopt the Modified Booth Encoding (MBE)\cite{8_farooqui1998general,14_kang2004fast,15_kang1993design,26_yeh2000high} for the first step (Fig.\ref{booth}), as it can reduce the number of partial product rows for an $n$-bit fixed-point multiplication by half to $n/2$, significantly decreasing the latency and hardware cost. This involves encoding the multiplicand $A$ and using the encoded $A$ and the multiplier $B$ to generate partial products. Next, in the second step, methods such as Wallace Tree\cite{24_wallace1964suggestion} or Compressor Tree\cite{19_santoro1989spim} are used to compress the partial products, effectively reducing the number of partial product rows to the final two rows (sums and carries). In the third step, designers often employ advanced adder technologies, such as carry-lookahead adders or carry-select adders \cite{18_hennessy2011computer}, to merge these two rows, thereby yielding the final product result.

Considering a single multiplication operation, we can draw two conclusions:  first, the encoding part is a logical computation related only to the multiplicand $A$ and is independent of the multiplier $B$; second, the result of the multiplication is directly related to the encoded multiplicand $A$ and the multiplier $B$, and indirectly related to the origin $A$.

\begin{figure} [htbp]
  \centering 
  \includegraphics[scale=0.5]{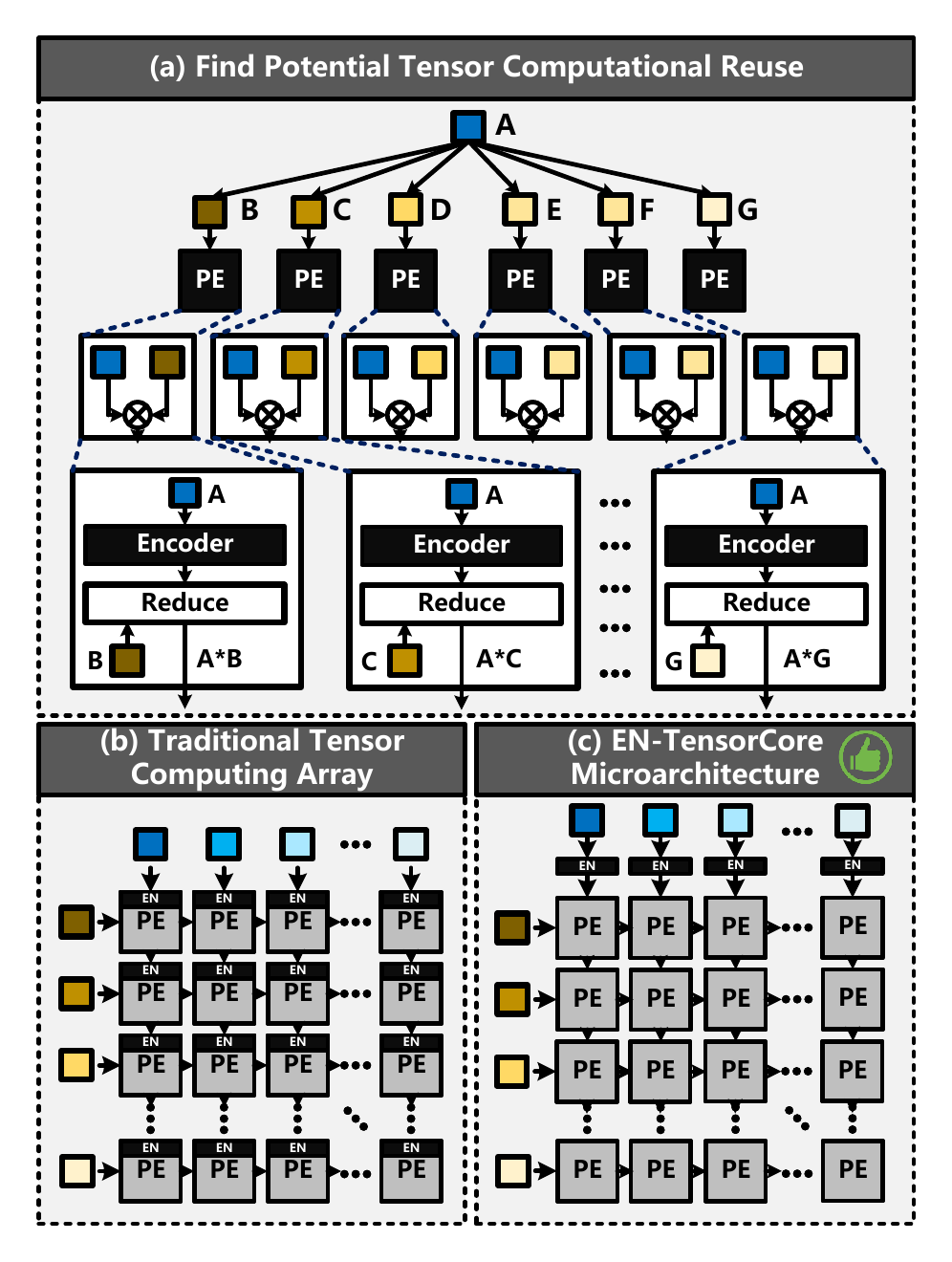}
   \vspace{-0.4cm}
  \caption{(a) The internal computational abstraction of PE. (b) From the perspective of TCU. (c) The proposed architecture.}
   \vspace{-0.55cm}
  \label{en_tensor}

\end{figure}

When we extend this behavior to arrays of multipliers, since matrix multiplication or convolution has phenomenon of same multiplicand by different multiplier in the spatial dimension or time dimension, there is a repeated encoding behavior of the multiplicand $A$ inside the multiplier of PEs (Fig.\ref{en_tensor}(a)). From the perspective of TCUs, what is needed is the encoded multiplicand $A$, not $A$ itself. When applied to the existing various TCUs hardware architectures, whether it's based on data broadcasting like 2D Matrix or 1D/2D Array, or data flow-based like Systolic Array or 3D Cube, enhancing performance only requires the following simple steps: First, remove the encoder logic from all 
multipliers within the tensor cores (Fig.\ref{en_tensor}(b)), 
retaining only the partial product compressor and the full adder; second, add a single encoder to each column of the multiplicand pathway outside the array (Fig.\ref{en_tensor}(c)), allowing the encoded multiplicand $A$ to flow or broadcast within the array. In terms of area, this can reduce the size of an individual PE, and when applied to TCUs with large-scale multiplier arrays, it can make the array layout more efficient and compact, which is beneficial for reducing the latency of multiplication operations. 
% In terms of power consumption, the approach achieves efficiency in two key ways. First, it reduces power consumption by avoiding repetitive encoding logic. Second, by decreasing the array area, it shortens the data transmission pathways between adjacent processing elements (PEs). This not only leads to direct energy savings but also minimizes the power needed for data flow between PEs. (such as the transfer of partial results and operands in Systolic Array (WS) and 3D cube, and the transfer of operands in Systolic Array (OS)).

In terms of power consumption, it not only saves the power consumption caused by repetitive encoding logic, but also, due to the reduced array area, makes the data transmission pathways between adjacent PEs shorter, which further reduces the power consumption associated with data flow between PEs.

\subsection{Challenge in Modified Booth Encoding}

Considering the multiplication of two $n$-bit integer $A$ (multiplicand) and $B$ (multiplier) in 2’s complement representation, i.e.,
\begin{equation}
    \begin{array}{l}
    A=-a_{n-1} 2^{n-1} +\sum _{i=0}^{n-2} a_{i} 2^{i}\\
    B=-b_{n-1} 2^{n-1} +\sum _{i=0}^{n-2} b_{i} 2^{i}\\
    \end{array}
\end{equation}

in MBE, $A$ is transformed into:
\begin{equation}
\label{encodingA}
    A=\sum _{i=0}^{\frac{n}{2} -1} m_{i} 2^{2i} =\sum _{i=0}^{\frac{n}{2} -1}( -2a_{2i+1} +a_{2i} +a_{2i-1}) 2^{2i}
\end{equation}

where $a_{-1} =0$, and $m_{i} \in \{-2,-1,0,1,2\}$.

Based on the encoding result of $A$ (the logical expression as in Eq. \ref{logic}), Booth selectors choose $-2B$, $-B$, $0$, $B$, or $2B$ to generate the partial product rows of $A\times B$ (Fig.\ref{booth}(a)). For a single-cycle multiplier, $\lceil n/2 \rceil$ encoders are required to encode the multiplicand in parallel (Fig.\ref{booth}(b)).
\begin{equation}
\label{logic}
\begin{array}{l}
NEG=\ a_{2i+1}\left(\overline{a_{2i}} +\overline{a_{2i-1}}\right)\\
SE=\overline{\overline{a_{2i+1}} a_{2i} a_{2i-1} +a_{2i+1}\overline{a_{2i}}\overline{a_{2i-1}}}\\
CE=\overline{a_{2i} \oplus a_{2i-1} +SE}\\
\end{array}
\end{equation}
% \begin{figure*} [htbp]
%   \centering 
%   \includegraphics[scale=0.35]{eq.pdf}
%   \caption{The Particular Solution of the Nonlinear Equation W}
%   \label{eq}
% \end{figure*}
MBE can be viewed as digit-set conversion: the recoding takes a radix-4 number with digits in $[0, 3]$ and converts it to the digit set $[-2, 2]$. The digit-set conversion process defined by radix-4 Booth’s recoding entails no carry propagation. Each radix-4 digit in $[-2, 2]$ is obtained, independently from all others, by examining 3 bits of the multiplicand, with consecutive 3-bit segments overlapping in 1 bit. Thus, radix-4 Booth’s recoding is said to be based on overlapped 3-bit scanning of the multiplicand.

However, applying MBE to the EN-T architecture does not achieve the expected effect. This is because every 2 bits of the multiplicand need to be encoded into 3 bits to serve as control lines NEG, SE, CE. Thus, for the multiplication of two $n$-bit numbers, the multiplicand needs to be encoded into $\lceil n/2 \rceil*3$ bits. Externalizing the encoder would actually cause the width of the interconnects between multiplicands in the TCUs, which is undesirable. The increase in interconnect width significantly affects the chip's area and power. To address this, we redesigned an encoder that encodes an $n$-bit multiplicand into $n+1$ bits to reduce the width of the encoded numbers, which will be described in the next subsection.

\subsection{Modified Encoding in EN-T architecture}
\subsubsection{Construction of Encoding Polynomials}

From Eq. \ref{encodingA}, it can be found that the intrinsic reason for the high bit width of the MBE is that the encoding coefficients $m_i$ have five different states $\{-2,-1,0,1,2\}$, therefore requiring 3 bits of control lines to select the corresponding multiplier. %multiplier同时有乘数/乘法器意思，此处容易歧义 
One approach is to start by compressing the number of states in the coefficients of the powers. We consider this issue from the perspective of the number decomposition; the partial product of $A\times B$ depends on the number of terms $A$ is decomposed into. A number with $n$ bits (we assume that $n$ is an even integer) can be decomposed into $\lceil n/2 \rceil$ terms by MBE, which means that the composition of each term must include powers of 4. Thus, an $n$-bit unsigned number can be decomposed into the following polynomial, where $a_i$ is a 2-bit unsigned number, and $a_i \in \{0,1,2,3\}$, $N=\frac{n}{2}$, $Q_N\in[0,2^n-1]$.
\begin{equation}
    Q_N=\sum _{i=0}^{N} a_{i} 4^{i}
\end{equation}

$Q_N\in[0,2^{n-1} + (2^{n-2} -1)]$ when $ a_{N+1} \in \{0,1,2\} $, $ a_i \in \{0,1,2,3\} $.

We aim to avoid the occurrence of 3 in the coefficients of the polynomial, as we cannot directly obtain the product of 3 and the multiplier B through shift operations. Consequently, we have restructured the coefficients of the polynomial $Q'$ as Eq. \ref{Q'}, where ${Cin}_{n+1} \in \{0, 1\}$.
% $w_i$ is a 2-bit encoded integer, and $ Encode(w_i)\in \{2'b00, 2'b01, 2'b10, 2'b11\}$ denote $\{0, 1, 2, -1\}$ respectively.
For $w_i \in \{0, 1, 2, -1\}$, we encode it in binary format and correspondingly map it to $\{00, 01, 10, 11\}$. We denote the encoding result as $Encode(w)$.
\begin{equation}
\label{Q'}
    Q'_N={Cin}_{N+1}4^{N+1} + \sum _{i=0}^{N} w_{i} 4^{i}
\end{equation}

If we can express Q in the form of Q', then we can transform the multiplication operation into several simple shift and addition operations. The current issue is whether we can find such a set of solutions $\{{Cin}_{N+1}, w_N, w_{N-1}, ..., w_0\}$ that satisfies $Q' = Q$. We will use mathematical induction to prove the existence of such solutions and obtain their recursive expression.

When $n = 0$, we have the following equation:
\begin{equation}
\label{eq:q0}
    Q_0 = a_0, \quad\quad Q'_0 = Cin_1 4 + w_0
\end{equation}

The equation $Q_0 = Q'_0$ holds true when we let:
\begin{equation}
\label{eq:w0}
w_0=\begin{cases}
 a_0, & a_0 \in \{0, 1, 2\} \\
 -1, & a_0 = 3
\end{cases}
,\quad
Cin_1=\begin{cases}
 0, & a_0 \in \{0, 1, 2\} \\
 1, & a_0 = 3
\end{cases}
\end{equation}
\begin{figure}[htbp]\centering\includegraphics[scale=0.5]{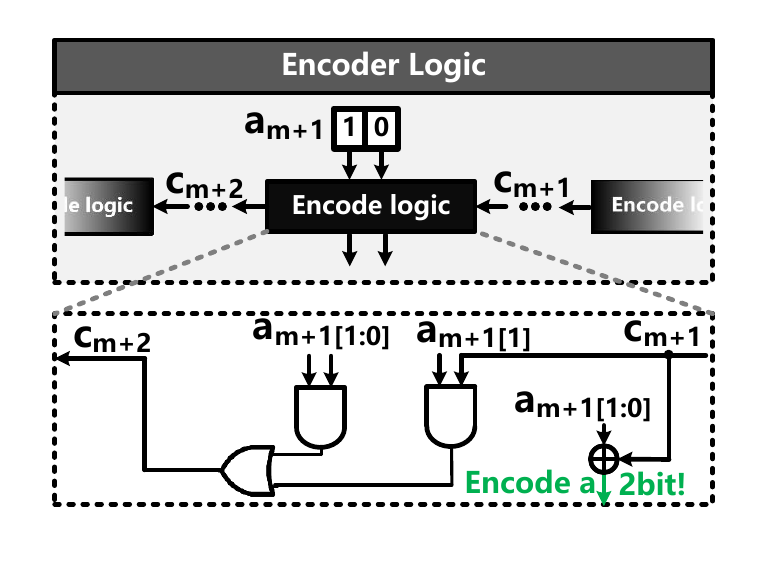}
   \vspace{-0.5cm}
\caption{Modified encoder logic.}
\label{cc}
   \vspace{-0.5cm}
\end{figure}
At this point, the binary encoding of the signed number $w_0$ and the 2-bit unsigned number $a_0$ are identical. This can be further expressed as:
\begin{equation}
\label{eq:enw0}
    Encode(w_0)=\left[a_0\right]_2, \quad Cin_1 = a_0\left[1\right] \& \,a_0\left[0\right]
\end{equation}
where $\left[a\right]_2$ denotes the binary encoding of the number $a$, and $a\left[i\right]$ represents the $i\emph{th}$ bit of $a$ in its binary encoding.

When $n = 1$, we have:
\begin{equation}
\label{eq:q1}
    Q_1 = a_1  4 + a_0, \quad Q'_1 = Cin_2 4^2 + w_1  4 + w_0
\end{equation}

Substituting Eq. \ref{eq:q0} into Eq. \ref{eq:q1} results in:
\begin{equation}
\begin{aligned}
Q_1 &= (a_1+Cin_1)  4 + w_0 \\
    &= a'_1  4 + w_0
\end{aligned}
\end{equation}
where $a'_1 = (a_1 + Cin_1) \in \{0,1,2,3,4\}$.

Letting:
\begin{equation}
\label{eq:w1}
w_1=\begin{cases}
 a_1', & a_1' \in \{0, 1, 2\} \\
 a_1' - 4, & a_1' \in {3, 4}
\end{cases}
,\quad
Cin_2=\begin{cases}
 0, & a_1' \in \{0, 1, 2\} \\
 1, & a_1' \in {3, 4}
\end{cases}
\end{equation}

Then $Q_1 = Q'_1$, and $w_1$ and $Cin_2$ can be expressed as:
\begin{equation}
\label{eq:enw1}
\begin{aligned}
&Encode(w_1) =\left[a_1\right]_2 + Cin_1 \\
&Cin_2 = \left(a_1\left[1\right] \& \, a_1\left[0\right]\right) | \left(a_1\left[1\right] \& \,Cin_1\right)  
\end{aligned} 
\end{equation}

Assume $Q_m = Q'_m$ holds true, this implies:
% \begin{equation}
% \label{eq:qn}
%     a_m4^m + a_{m-1}4^{m-1} + ... + a_0 = Cin_{m+1}4^{m+1} + w_m4^m + w_{m-1}4^{m-1} + ... + w_0
% \end{equation}
\begin{equation}
\label{eq:qn}
    \sum_{i=0}^{m} a_i 4^i = Cin_{m+1}4^{m+1} + \sum_{i=0}^{m} w_i 4^i
\end{equation}

To ensure that $Q_{m+1} = Q'_{m+1}$, it is equivalent to proving $Q_{m+1} - Q_m = Q'_{m+1} - Q'_m$, which can be expressed as:
\begin{equation}
\label{eq:qn+1}
    (a_{m+1} + Cin_{m+1})4^{m+1} = Cin_{m+2}4^{m+2} + w_{m+1}4^{m+1}
\end{equation}
This can be simplified to:
\begin{equation}
\label{eq:qn+1'}
    a'_{m+1}= Cin_{m+2} 4 + w_{m+1}
\end{equation}
where $a'_{m+1} = a_{m+1} + Cin_{m+1}$. Similar to the values in Eq. \ref{eq:w1} and the encoding method in Eq. \ref{eq:enw1}, we can select the following expression to ensure Eq. \ref{eq:qn+1'} holds true:
\begin{equation}
    \label{eq:wn+1}
\begin{aligned}
&w_{m+1}=\begin{cases}
 a'_{m+1}, & a'_{m+1} \in \{0, 1, 2\} \\
 a'_{m+1} - 4, & a'_{m+1} \in \{3, 4\}
 \end{cases}
% \end{equation}
% \begin{equation}
% \label{eq:wn+2}
\\&Cin_{m+2}=\begin{cases}
 0, & a'_{m+1} \in \{0, 1, 2\} \\
 1, & a'_{m+1} \in \{3, 4\}
\end{cases}
\end{aligned}
\end{equation}

\begin{equation}
\label{eq:enwn+1}
\begin{aligned}
&Encode(w_{m+1}) =\left[a_{m+1}\right]_2 + Cin_{m+1} \\
&Cin_{m+2} = \left(a_{m+1}\left[1\right] \&\, a_{m+1}\left[0\right]\right) | \left(a_{m+1}\left[1\right] \& \,Cin_{m+1}\right)  
\end{aligned} 
\end{equation}

Based on the foregoing analysis, we have demonstrated that for $n=2$ and $n=m$, through the recursive expressions Eq. \ref{eq:wn+1} and Eq. \ref{eq:enwn+1}, a set of solutions $\{Cin_{N+1}, w_N, w_{N-1}, ..., w_0\}$ can be identified, enabling the polynomial $Q_N$ to be expressed as the polynomial $Q'_N$. According to the principle of mathematical induction, for all $N \geq 2$, we can employ a recursive method to represent the polynomial $Q_N$ as the polynomial $Q'_N$. Hence, based on Eq. \ref{eq:w0}, \ref{eq:enw0}, \ref{eq:wn+1} and \ref{eq:enwn+1}, we can encode an $n$-bit unsigned number into $n/2$ two-bit coefficients and one one-bit coefficient.
Modified encoding logic as show in Fig.\ref{cc}.

% Therefore, when $n\geq 2$, the numerical representation range of $W$ can always include $[0,2^{n-1}]$. Thus, we can encode the $n$-bit signed number $A$ as $Encode(A) = \{Sign, w_{N+1}, w_N, ..., w_1\}$, with each item except the sign bit represented by 2 bits, making the total data width $n+1$ bits.

\begin{figure*} [htbp]
  \centering 
  \includegraphics[scale=0.215]{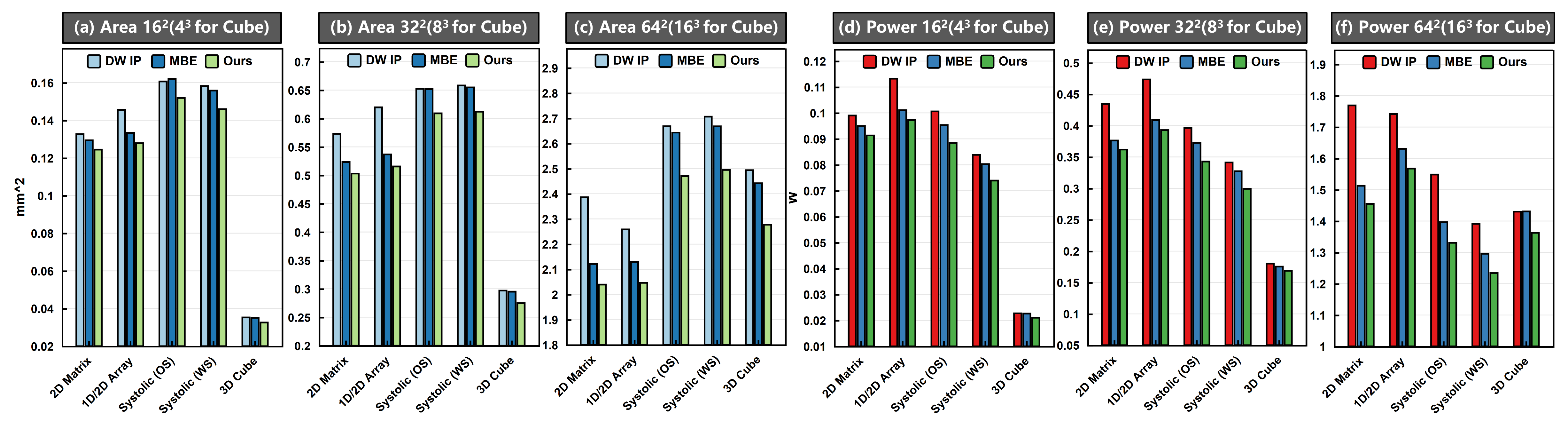}
    \vspace{-0.4cm}
  \caption{Comparison of area and power consumption under different TCU architectures and array sizes.}
  \label{area_power}
    \vspace{-0.4cm}
\end{figure*}

\subsubsection{Computing Method}
For example, 78 (INT8) is encoded as $Encode(78)=\color{cyan}\{0,1,1,-1,2\}$, where the first number is the sign, and the next four 2-bit numbers are the encoding numbers, therefore the multiplication result of $B$ and 78 is ${\color{cyan}1}*B4^3+{\color{cyan}(1)}*B4^2+{\color{cyan}(-1)}*B4^1+{\color{cyan}(2)}*B$. The subsequent calculations are completely identical to those of MBE, that is, $B$ with different bit weights is added together to obtain the final multiplication result. When $A$ is negative, it is only necessary to perform a transformation on $B$ based on the sign bit of $A$ to implement signed multiplication (at this point, the hardware will choose $-B$ as the $B$ in the above formula to participate in the actual calculation). 
% After theoretical analysis of feasibility, the next step is to design the encoder, which involves obtaining the specific value of $\{w_{N+1},w_N,...,w_1\}$ from the multiplicand $A$. This is essentially a problem of solving a nonlinear equation. In the next part, we will use mathematical induction to obtain the exact analytical solution and analyze the logical recursive expression.

In terms of the number of encoders, for an $n$-bit signed number using MBE, $\lceil n/2 \rceil$ encoders are required; we need $\lceil n/2 \rceil-1$ encoders (as the lowest 2 bits do not need encoding). For the encoded bit width, MBE requires $\lceil n/2 \rceil*3$ bits, whereas our method requires $n+1$ bits. In the experimental section, we will first perform an evaluation of the area, delay, and other performance metrics of encoders for different bit widths, and test the performance of the optimized encoders within the EN-T architecture.

\section{Experiment}
\subsection{Implement Environment}
We implement our design in RTL and then synthesize, place and route it with Synopsys Design Compiler toolchain using the SMIC 40nm NLL-HS-RVT technology and  using ARM Memory Compiler to generate on-chip SRAM. We evaluate the performance and energy costs using PrimeTime PX based on VCD waveform files obtained from simulation with typical corner process.

\begin{table}[]
\setlength{\tabcolsep}{2mm}{
        \renewcommand{\arraystretch}{0.9}
\begin{tabular}{ccccccc}

\midrule[0.7pt]\midrule[0.7pt]
\multicolumn{7}{c}{Single Encoder Comparison}                                                                                                                                  \\ 
\midrule
\multicolumn{2}{c}{Method}                       & AND                   & NAND                  & NOR                   & XNOR                    & Area                      \\ 
\midrule
\multicolumn{2}{c}{MBE}                          & 2                     & 2                     & 1                     & 1                       & 7.06                     \\
\multicolumn{2}{c}{Ours}                         & 1                     & 3                     & 0                     & 2                       & 8.64                     \\ \midrule[0.7pt]\midrule[0.7pt]
\multicolumn{7}{c}{Comparison of High Bit Encoders}                                                                                                                            \\ 
\midrule
Width                  & Method                  & Area                  & Delay                 & Power                 & Number                  & En-Width                  \\ 
\midrule
\multirow{2}{*}{8}     & MBE                     & 28.22                & 0.23                  & 24.06                 & 4                       & 12                        \\
                       & Ours                    & 25.93               & 0.36                  & 21.47                 & 3                       & 9                         \\ 
                       \midrule[0.2pt]
\multirow{2}{*}{10}    & MBE                     & 35.28                 & 0.23                  & 30.07                 & 5                       & 15                        \\
                       & Ours                    & 34.57               & 0.45                  & 28.47                 & 4                       & 11                        \\ 
                       \midrule[0.2pt]
\multirow{2}{*}{12}    & MBE                     & 42.34                & 0.23                  & 36.03                 & 6                       & 18                        \\
                       & Ours                    & 42.22                & 0.54                  & 35.49                 & 5                       & 13                        \\ 
                       \midrule[0.2pt]
\multirow{2}{*}{14}    & MBE                     & 49.39                & 0.23                  & 42.03                 & 7                       & 21                        \\
                       & Ours                    & 50.86               & 0.63                  & 42.45                 & 6                       & 15                        \\ 
                       \midrule[0.2pt]
\multirow{2}{*}{16}    & MBE                     & 56.45                & 0.23                  & 48.05                 & 8                       & 24                        \\
                       & Ours                    & 60.51               & 0.71                  & 49.40                 & 7                       & 17                        \\ 
                       \midrule[0.2pt]
\multirow{2}{*}{18}    & MBE                     & 63.50                & 0.23                  & 54.01                 & 9                       & 27                        \\
                       & Ours                    & 69.15               & 0.80                  & 56.36                 & 8                       & 19                        \\ 
                       \midrule[0.2pt]
\multirow{2}{*}{20}    & MBE                     & 70.56                 & 0.23                  & 60.00                    & 10                      & 30                        \\
                       & Ours                    & 77.79               & 0.89                  & 63.38                 & 9                       & 21                        \\ 
                       \midrule[0.2pt]
\multirow{2}{*}{24}    & MBE                     & 84.67                & 0.23                  & 71.96                 & 12                      & 36                        \\
                       & Ours                    & 95.08               & 1.06                  & 77.23                 & 11                      & 25                        \\ 
                       \midrule[0.2pt]
\multirow{2}{*}{32}    & MBE                     & 112.90               & 0.23                  & 95.89                 & 16                      & 48                        \\
                       & Ours                    & 129.65               & 1.41                  & 105.14                & 15                      & 33                        \\ \midrule[0.7pt]\midrule[0.7pt]
\multicolumn{7}{c}{Multiplier Performance Comparison}                                                                                                                          \\ 
\midrule
bit                    & \multicolumn{3}{c}{Method}                                              & Area                  & Delay                   & Power                     \\ 
\midrule
\multirow{4}{*}{INT8}  & \multicolumn{3}{c}{DW IP}                                               & 291.6                 & 1.87                    & 211.4                     \\
                       & \multicolumn{3}{c}{MBE}                                                 & 292.7                 & 1.86                    & 212.2                     \\
                       & \multicolumn{3}{c}{Ours}                                                & 290.4                 & 1.99                    & 210.3                     \\
                       & \multicolumn{3}{c}{RME\_Ours}                                            & 264.4                & 1.63                    & 188.9                      \\ \midrule[0.7pt]\midrule[0.7pt]
\end{tabular}
}
\caption{Area/$\mu m^2$, Delay/$ns$ and Power/$\mu W$}
\label{table1}
  \vspace{-1cm}
\end{table}

\subsection{Performance of the Encoder}
In Table. \ref{table1}, we ran performance tests on encoders for both 2-bit and multi-bit cases. In the comparison of resource consumption for 2-bit encoders, our method requires one less AND gate but one additional XNOR gate compared to MBE. This is because the XOR logic is used to generate the sum for both 2 bits in Eq. \ref{eq:enwn+1}. Although our individual encoder is at a disadvantage in terms of area, in the case of multi-bit encoding tests, our encoding representation requires one less encoder due to the fact that the lowest 2 bits do not need encoding. Therefore, our method only exhibits advantages in terms of area and power consumption when the encoding bit width is less than 14 bits. In terms of latency, due to the parallel computation nature of MBE encoding, the extended bit width has almost no impact on its latency. However, our encoding is based on carry-chain encoding, causing the latency to gradually increase with the width of the multiplicand. This is a drawback of our method, but it becomes an insignificant factor in the EN-T architecture, as the encoders will be placed outside the array and enter the array through registers. At this point, the true critical path lies within the adders and accumulators inside the PE. Lastly, in terms of the comparison of encoding bit width, our encoding representation is not sensitive to the data bit width, whereas MBE's data bit width increases by a factor of 1.5. From the perspective of a single multiplier, this may not be an important factor. However, in the EN-T architecture, it becomes an obstacle that limits the performance improvement of the array. We also conducted performance tests on multipliers, using Synopsys DesignWare standard process library multiplier (DW IP) and Modified Booth Multiplier as benchmarks. In terms of INT8 performance, the area and power consumption are comparable to those of DW IP, with a slightly higher delay of 0.12ns. However, this is not the primary factor. We designed it specifically for the EN-T architecture. In the experiments after removing the encoder logic (RME\_Ours), there are significant improvements in area, delay, and power consumption, making it a promising solution for achieving significant performance improvements in large-scale TCUs.

\begin{figure*} [htbp]
  \centering 
  \includegraphics[scale=0.212]{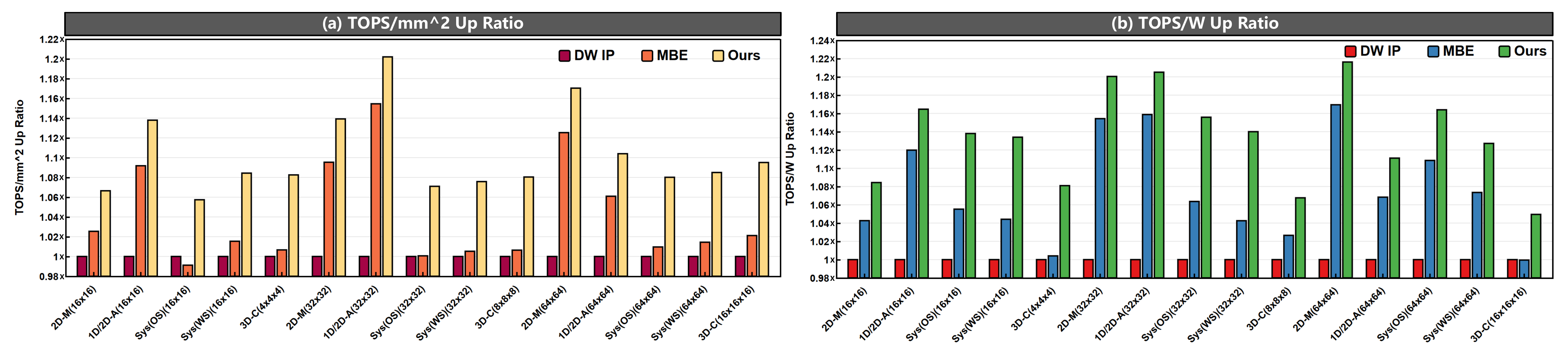}
  \vspace{-0.3cm}
  \caption{Up ratio of (a) area efficiency and (b) energy efficiency.}
  \label{upratio}
    \vspace{-0.3cm}
\end{figure*}

\subsection{Performance of EN-T architecture}
% 我们的测试范围覆盖主流AI加速器的几种张量计算微架构2D-Matrix(Fig.2(a))，1D/2D Array(Fig.2(b)),两种类型的Systolic Array(WSand OS) (Fig.2(c)(d)),和3D Cube(Fig.2(e)),阵列规模将在16^2,32^2,64^2(Cube为4^3,8^3,16^3)下测试EN-T architecture架构(Fig.3(c))的可扩放性，综合评估能效，面效等参数。测试基准对象为Synosys DesignWare Standard process library multiplier构成的PE，EN-T architecture架构采用2种编码器(Modified Booth Encoder和Ours Encoder)采用寄存器输出。PE中乘法器的精度为INT8，累加器位宽为16+

% ,S为阵列尺寸。
We did performance tests on EN-T architecture within the scope of mainstream AI accelerators, covering various tensor computation microarchitectures: 2D Matrix (Fig.\ref{architecture}(a)), 1D/2D Array (Fig.\ref{architecture}(b)), two types of Systolic Array (WS and OS) (Fig.\ref{architecture}(c)(d)), and 3D Cube (Fig.\ref{architecture}(e)). The array sizes will be tested at $16^2$, $32^2$, $64^2$ (Cube: $4^3$, $8^3$, $16^3$) to evaluate the scalability, energy efficiency, and area efficiency of the EN-T architecture (Fig.\ref{en_tensor}(c)). The benchmark object for testing is the PE composed of Synopsys DesignWare standard process library multiplier. The EN-T architecture utilizes two encoders (MBE and Our Encoder) with register outputs, and all test on 500MHz. The accuracy of multipliers in the PE is INT8 (Due to the prevalent use of INT8 in existing AI accelerators and edge devices, where the mantissa part of floating-point numbers often involves low-bit-width unsigned multiplication (BF16(UINT7), TF32(UINT10)), we adopt INT8 as the multiplication bit width for testing TCUs.), and the accumulator width is $16+\log_2 S$. $S$ represents the array size.

\begin{figure}[htbp]
\centering\includegraphics[scale=0.22]{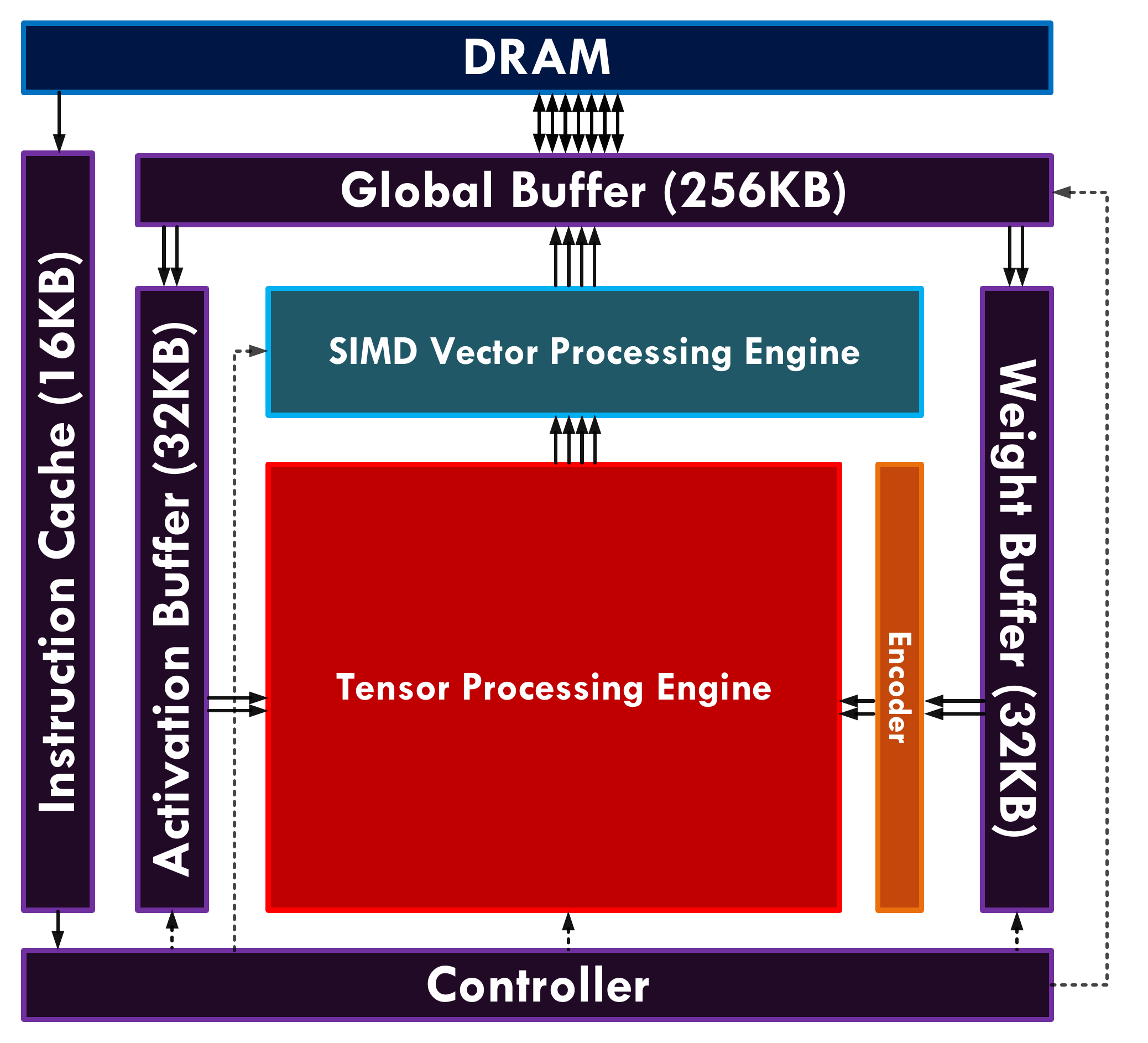}
  % \vspace{-0.2cm}
\caption{Benchmark SoC.}
\label{SoC}
  \vspace{-0.5cm}
\end{figure}

In the area test (Fig.\ref{area_power}(a)(b)(c)), EN-T architecture with MBE encoder demonstrates sensitivity to architectures. Even though the area of $S^2$ encoders is removed, the reduction in area is not significant in TCUs based on pipelined transfer, such as Systolic Array and 3D Cube. In some cases, there may even be an increase in area. This is due to the high data bit width of MBE encoding, which incurs the cost of $S^2$ 4-bit registers for transferring this data in Systolic Array. However, in data broadcast-based 2D Matrix and 1D/2D Array, there is no such area overhead, and the removed logic can compensate for the impact of MBE's extra layout wire width. On the other hand, our approach further compresses the data line width in EN-T, enabling its relative advantages over MBE in TCUs based on pipelined transfer. Therefore, our encoding strategy can achieve significant area reduction in these architectures, making the array more compact and efficient. In the power test (Fig.\ref{area_power}(d)(e)(f)), both EN-T architecture with MBE encoder and our encoder achieved significant reductions compared to the baseline. This is different from the area test. The reason is that the power consumption of an MBE 8-bit encoder is 24.07$\mu W$, while the additional power consumption for transferring 4-bit registers is approximately 15.13$\mu W$. The reduction in area also leads to shorter paths for data transfer between PEs, which helps further reduce power consumption. On the other hand, our encoder-based EN-T architecture, benefiting from lower encoding bit width and area, can further reduce power from data transfer compared to MBE.

\begin{table}[]
\setlength{\tabcolsep}{2.7mm}{
        \renewcommand{\arraystretch}{0.8}
\begin{tabular}{ccccccc}

\midrule[0.7pt]\midrule[0.7pt]

\multicolumn{7}{c}{\textbf{Global Buffer}}                                                                                                                          \\ 
\midrule
Size($KB$)                    & \multicolumn{2}{c}{Area($\mu m^2$)}             & \multicolumn{2}{c}{Read Power($W$)}               & \multicolumn{2}{c}{Write Power($W$)}                 \\ 
\midrule
\multirow{1}{*}{256}  & \multicolumn{2}{c}{614400}                                               & \multicolumn{2}{c}{0.0205}                 & \multicolumn{2}{c}{0.04515}                   \\ 
\midrule[0.7pt]\midrule[0.7pt]
\multicolumn{7}{c}{\textbf{Activation and Weight Buffer}}                                                                                                                          \\ 
\midrule
Size($KB$)                    & \multicolumn{2}{c}{Area($\mu m^2$)}             & \multicolumn{2}{c}{Read Power($W$)}               & \multicolumn{2}{c}{Write Power($W$)}                 \\ 
\midrule
\multirow{1}{*}{64}  & \multicolumn{2}{c}{153600}                                               & \multicolumn{2}{c}{0.0146}                 & \multicolumn{2}{c}{0.0322}                   \\ 
\midrule[0.7pt]\midrule[0.7pt]
\multicolumn{7}{c}{\textbf{SIMD Vector Processing Engine}}                                                                                                                          \\ 
\midrule
ALU               & \multicolumn{2}{c}{Precision}       & \multicolumn{2}{c}{Area($\mu m^2$)}             
& \multicolumn{2}{c}{Power($W$)}     \\ 
\midrule
\multirow{1}{*}{32}  & \multicolumn{2}{c}{TF32}                                               & \multicolumn{2}{c}{126481}                 & \multicolumn{2}{c}{0.0951}                   \\ 
\midrule[0.7pt]\midrule[0.7pt]
% \multicolumn{7}{c}{\textbf{Controller and Img2col}}                                                                                                                          \\ 
% \midrule
%  & Number  & \multicolumn{2}{c}{Area($\mu m^2$)}   & \multicolumn{2}{c}{Power($W$)}     \\ 
% \midrule
% & 2 & \multicolumn{2}{c}{83679}                    & \multicolumn{2}{c}{0.0632}                   \\ 
% \midrule[0.7pt]\midrule[0.7pt]
\end{tabular}
}
\caption{On-chip Parameters of the SoC Benchmark}
\label{table2}
\vspace{-0.7cm}
\end{table}

\begin{figure*}[htbp]\centering\includegraphics[scale=0.085]{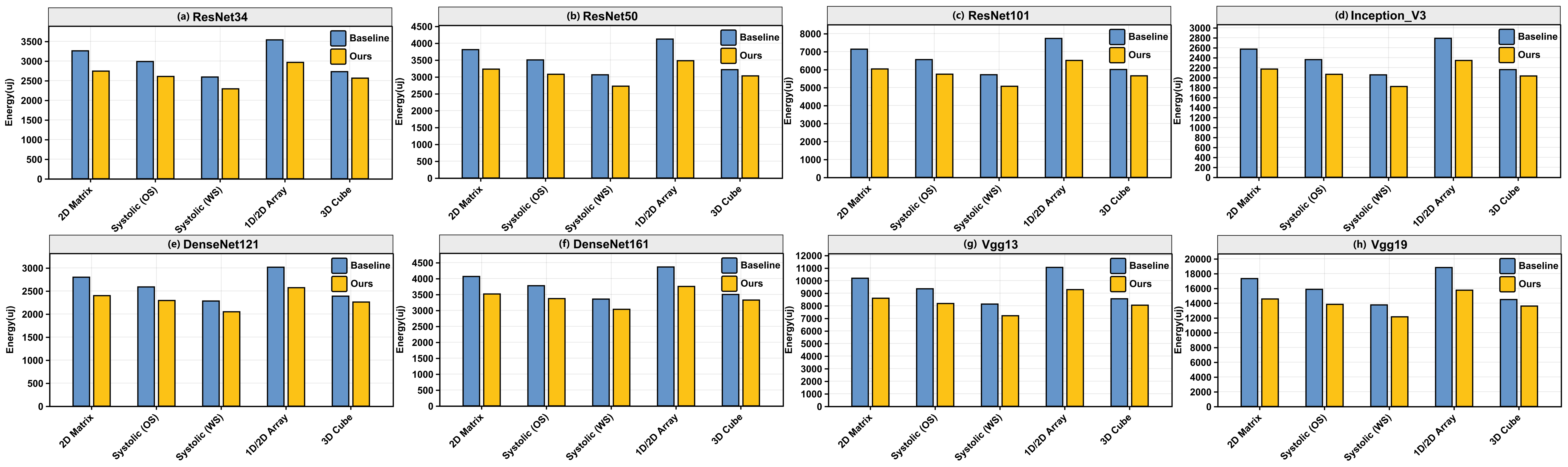}\caption{Single frame energy consumption comparison for SoC inference based on EN-T architecture.}
\label{eng}
  \vspace{-0.3cm}
\end{figure*}

\begin{figure*}[htbp]\centering\includegraphics[scale=0.15]{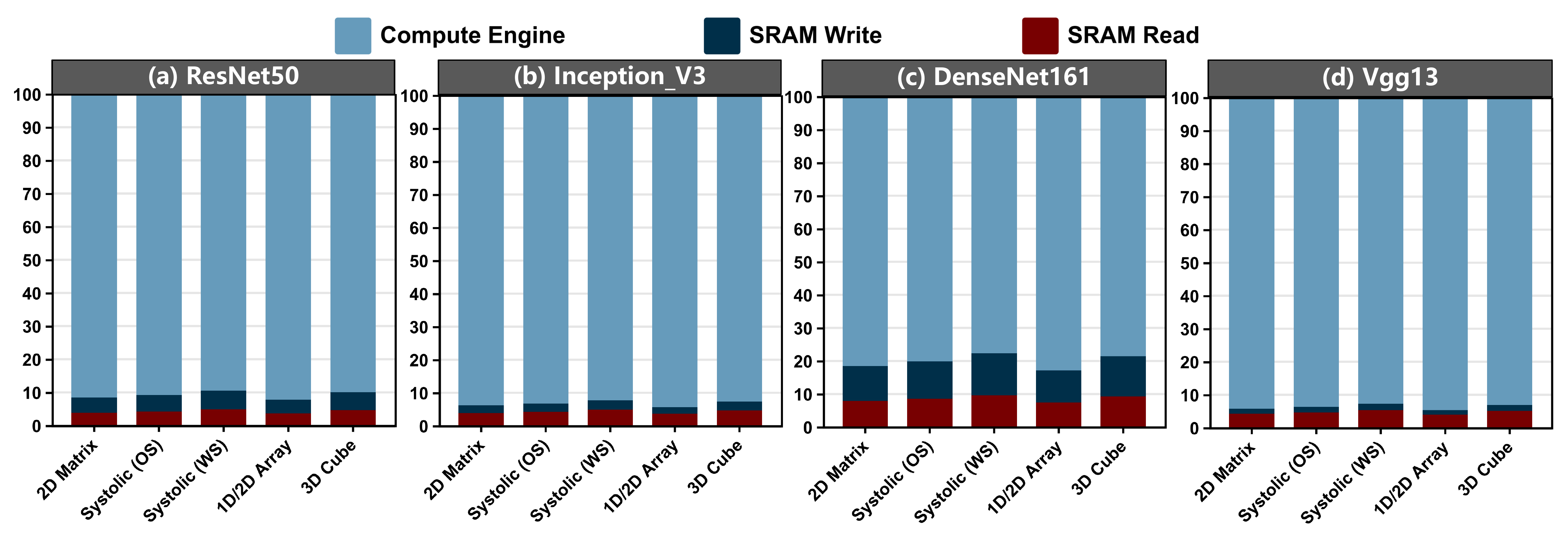}
  \vspace{-0.2cm}
\caption{SoC normalized energy fraction under baseline-based TCU.}
\label{ocu}
  \vspace{-0.3cm}
\end{figure*}

%在面积测试(Fig.5(a)(b)(c))中采用MBE编码器的EN-T architecture表现出对架构敏感，即使移除了S^2个编码器的面积，但在基于流水线传递的TCUs中例如SystolicArray与3D Cube面积降低并不明显，甚至个别情况会出现面积增加的现象，这是由于MBE编码数据位宽较高的原因，会在Systolic Array中对传递这部分数据付出S^2个4bit寄存器的代价，而在基于数据广播的2D Matrix与1D/2D Array则没有这部分面积开销，移除的逻辑可以补偿MBE多的布局线宽造成的影响。而基于我们的方案在EN-T architecture中进一步压缩了数据线位宽，使得在基于流水线传递的TCUs中能够展现出相对于MBE的优势，因此我们的编码策略在这几种架构下都能获得明显的面积降低，使得阵列更加紧凑与高效。在功耗测试(Fig.5(d)(e)(f))中采用MBE编码器的EN-T architecture与ours encoder都能获得相对于基准明显的降低，这是与面积测试不同的地方。原因是由于一个MBE 8bit编码器的功耗是24.07uw而多出4bit的寄存器传输功耗约为15.13uw。面积的降低又会使得PE之间传递数据的路径更短，有利于进一步降低功耗。而ours encoder basedEN-T architecture得益于编码位宽与面积更低的原因，使得相较于MBE能够进一步降低数据传输产生的功耗。

In the tests of energy efficiency and area efficiency (Fig.\ref{upratio}(a)(b)), we mainly compare the scalability of individual TCUs with different computational scales under EN-T architecture. Due to the square relationship between the number of removed encoders and the array size, increasing the computational scale to a certain extent will result in higher performance. As shown in Fig.\ref{upratio}(a)(b), when the computational scale of our encoder-based EN-T architecture expands from 256GOPS to 1TOPS and 4TOPS, the average area efficiency improves from 8.7\% to 12.2\% and 11.0\%, and the energy efficiency improves from 13.0\% to 17.5\% and 15.5\%, respectively. Among them, the 1D/2D Array achieves a 20.2\% increase in area efficiency and a 20.5\% increase in energy efficiency compared to the baseline at 1TOPS. This is due to the specific characteristics of the multiplier-adder architecture itself (with no PEs, multipliers and multiplicands are not pipelined to the adder tree). In this case, the performance improvement of the EN-T architecture is the highest.

\subsection{Benchmark on SoC}

In this section, we will evaluate the performance of the EN-T architecture on System-on-Chip (SoC) level for neural network inference (ResNet34, ResNet50, ResNet101, Inception\_V3, DenseNet121, DenseNet161, Vgg13, Vgg19). We use a basic NPU architecture and run on 500MHz, as shown in the Fig.\ref{SoC}, which includes three levels of storage. The SoC contains two levels of on-chip storage: a 256KB Global Buffer and 64KB Activation and Weight Buffers. In the readout of the Weight Buffer, we have added 32 Encoders to convert weights into encoded numbers for computation in the TCUs, with the Encoder module using register output. The Controller module includes control for reading and writing to SRAM and contains an img2col module for preprocessing the convolution operations. We use five types of TCU architectures: 2D Matrix, 1D/2D Array, two types of Systolic Array (WS and OS), and 3D Cube, with an array size of 32$\times$32. The 3D Cube configuration consists of two $8^3$ arrays, with a unified computational scale of 1024GOPS. Externally to the TCUs, there is a SIMD Vector Processing Engine, internally using 32 ALUs with TF32 precision for quantization, pooling, scalar addition, and activation functions. The detailed hardware parameters of each module are shown in Table.\ref{table2}.

\begin{figure*}[htbp]\centering\includegraphics[scale=0.1]{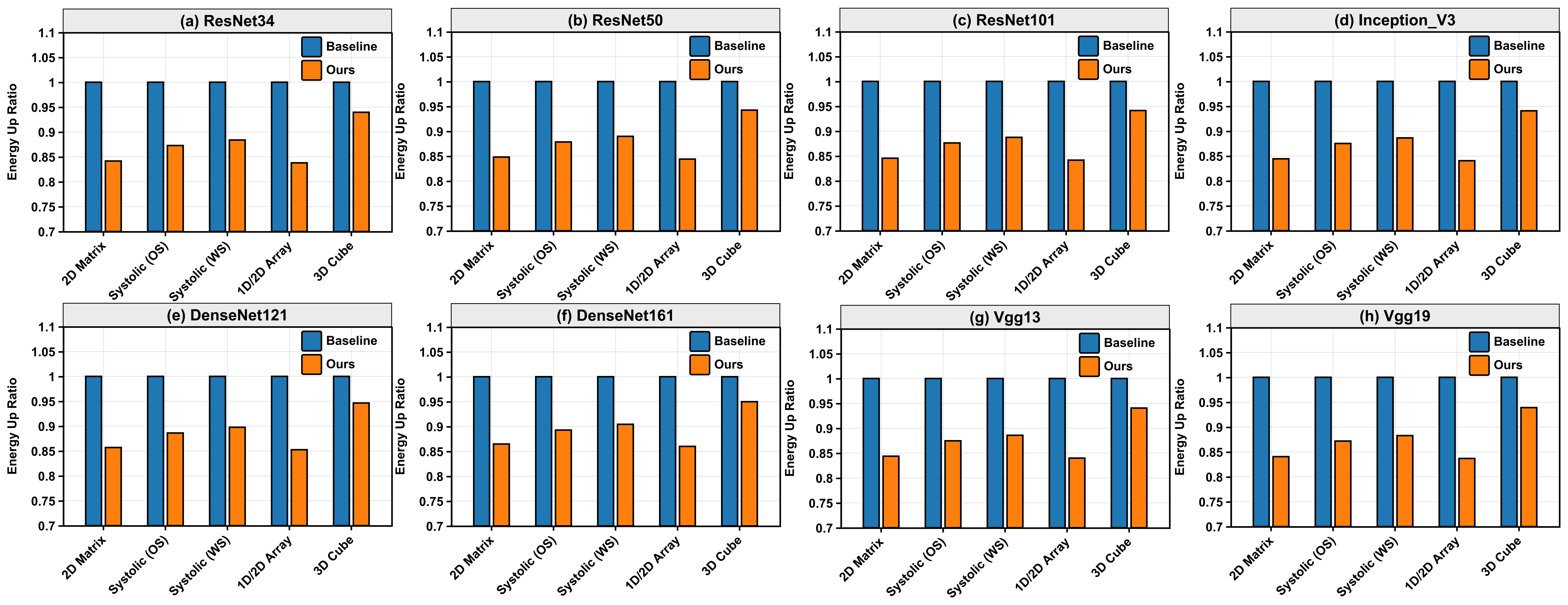}\caption{Energy consumption reduction ratio of SoC.}
\label{energy_3}
  \vspace{-0.3cm}
\end{figure*}

We conducted an energy consumption analysis for single-frame image (1,3,224,224)  inference on the entire SoC, starting with an analysis of the energy proportion normalized for the SoC based on the baseline TCU, as shown in Fig.\ref{ocu}. We decomposed the on-chip energy consumption into the read and write of SRAM, as well as the energy generated by the computing engines (TCU and SIMD Vector Processing Engine). For neural networks, the energy consumed by the computing engines accounts for a high proportion of the on-chip energy, ranging from 80-94\%. This is due to the high data reuse rate inherent in convolution and matrix multiplication. For instance, in a Systolic Array, each read and write operation to SRAM allows data to be transmitted and undergo multiply-accumulate operations across all rows and columns of the array, making the computing array become the primary source of power consumption. However, for some lightweight networks like DenseNet or Mobilenet that use a higher proportion of memory-intensive depthwise separable convolutions, the proportion of power consumption from memory access increases, but still does not exceed 25\% (as shown in Fig.\ref{ocu}(c)). Therefore, reducing the power consumption of the TCU is beneficial for lowering the system.

We replaced the baseline TCU with the EN-T architecture and extracted the encoder tests for all the adders to analyze the energy consumption before and after the replacement. The energy consumption of the SoC is shown in Fig.\ref{eng}, and the energy reduction ratio is shown in Fig.\ref{energy_3}. For the 2D Matrix architecture, the energy consumption can be reduced by 15.1\%-15.9\% across different networks; for the Systolic Array (OS) architecture, can be reduced by 11.3\%-12.8\%; for the Systolic Array (WS) architecture, can be reduced by 10.2\%-11.7\%; for the 1D/2D Array architecture, can be reduced by 14.0\%-16.0\%; for the 3D Cube architecture, can be reduced by 5.0\%-6.0\%. Among these, applying the EN-T architecture under the 3D Cube scenario yields lower benefits compared to other architectures. This is due to the 3D structure requiring more encoders for the same computational power. For example, a 32$\times$32 two-dimensional array requires 32 encoders, saving 992 encoders, while to achieve 1024 GOPS of computational power with a 3D Cube, two $8^3$ arrays are needed, requiring 128 encoders and saving 896 encoders. Therefore, the energy saving benefits are not as significant as those of other 2D architectures.

\begin{figure}[htbp]\centering\includegraphics[scale=0.14]{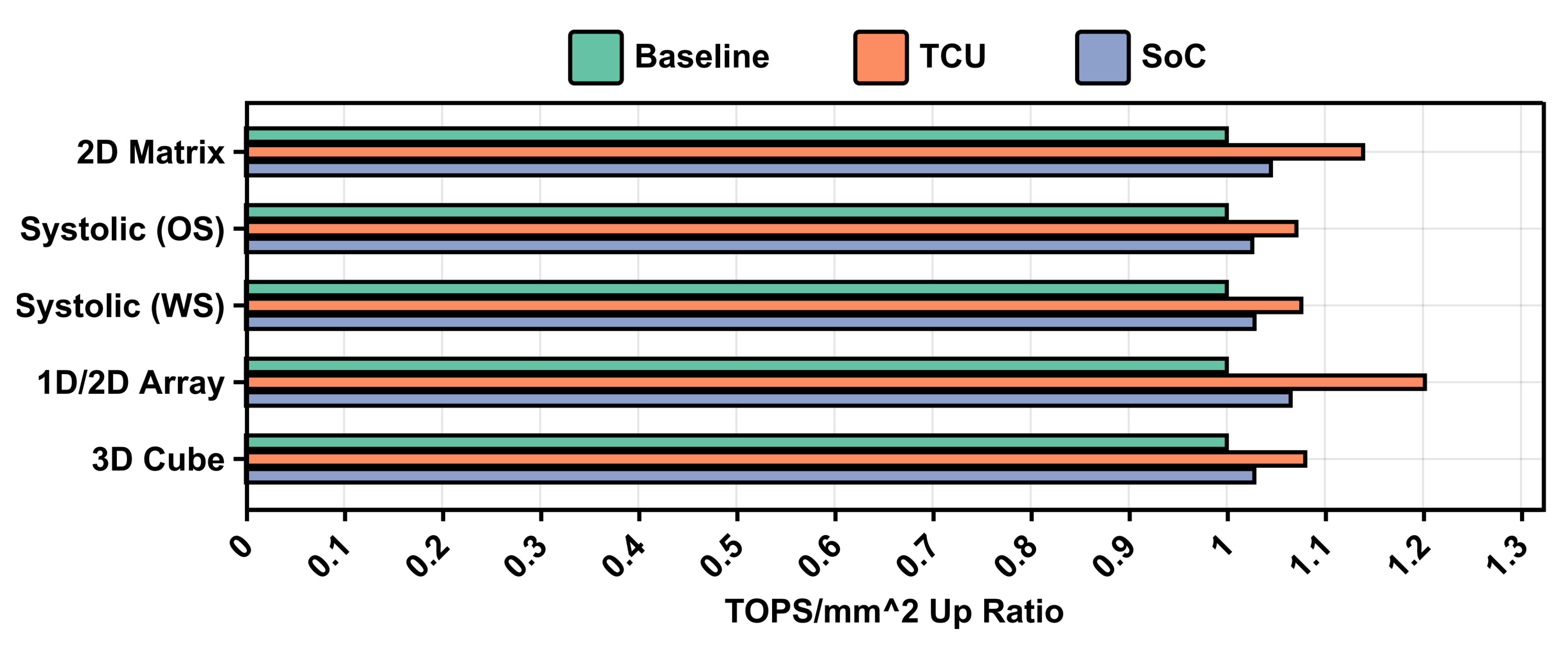}
  \vspace{-0.3cm}
\caption{Area efficiency of the SoC.}
\label{SoCmm}
  \vspace{-0.4cm}
\end{figure}

In terms of area efficiency, we compared the area efficiency improvement ratios of the baseline as show in Fig.\ref{SoCmm}, as well as the individual TCU and SoC. Due to the addition of on-chip SRAM, Controller, and SIMD Vector Processing Engine, as shown in Table.\ref{table2}, the area proportion of the on-chip SRAM is basically equivalent to that of the computing modules. Therefore, from the perspective of SoC, the area benefits brought by applying the EN-T architecture are relatively low. The main advantage is that it can reduce the inference power consumption by 10\%-16\%.

\section{Conclusion}
This paper explores new avenues in addressing the computational reuse issues in traditional TCUs by designing a matrix multiplication unit with an EN-T architecture. This method boasts high versatility and can be directly applied to existing tensor computing engines. From the perspective of computational encoding, we propose a method for encoding data with lower bit widths within the multiplier and uncover new opportunities for reducing energy consumption from the perspective of the multiplier's internals.

\bibliographystyle{IEEEtranS}
\bibliography{refs}

\end{document}